\documentclass[11pt]{JHEP3}
\usepackage{amsmath, amsthm,amsfonts,amssymb,amscd,citesort,mathrsfs,graphicx,fontenc,Glebdefs}

\usepackage[T1]{fontenc}

\author{
Marius de Leeuw$^a$\footnote{E-mail: marius.de.leeuw@aei.mpg.de} \, and\, Tomasz  {\L}ukowski$^b$\footnote{E-mail: lukowski@mathematik.hu-berlin.de} \\  
{\it$^a$Max-Planck-Institut f\"ur Gravitationsphysik\\
~Albert-Einstein-Institute\\
~Am M\"uhlenberg 1,\\
~14476, Potsdam, Germany\\

$^b$Institut f\"ur Mathematik und Institut f\"ur Physik, Humboldt-Universit\"at zu Berlin\\
Johann von Neumann-Haus, Rudower Chaussee 25,\\ 12489 Berlin, Germany}
}

\abstract{
In this paper we derive both the leading order finite size corrections for twist-2 and twist-3 operators and the next-to-leading order finite-size correction for twist-2 operators in beta-deformed SYM theory. The obtained results respect the principle of maximum transcendentality as well as reciprocity. We also find that both wrapping corrections go to zero in the large spin limit. Moreover, for twist-2 operators we studied the pole structure and compared it against leading BFKL predictions. 

}

\title{Twist operators in N=4 beta-deformed theory}

\preprint{
          \tiny{HU-Mathematik: 2010-21}\\[-.5ex]
          \tiny{HU-EP-10/86}\\[-.5ex]
          \tiny{AEI-2010-179}\\[-.5ex]
          }

\begin{document}
\renewcommand{\thefootnote}{\arabic{footnote}}
\setcounter{footnote}{0}

\section{Introduction}

The study of exact scaling dimensions of ${\cal N}=4$ SYM gauge-invariant composite operators with finite quantum numbers, through the gauge-string correspondence \cite{Maldacena:1997re}, has recently undergone significant developments coming from the use of integrability techniques. The finiteness of the system forces one to supplement the asymptotic result described by the Asymptotic Bethe Ansatz \cite{Beisert:2005fw}  with finite-size corrections described by the so-called L\"uscher formulae \cite{Luscher:1985dn}. In the context of the AdS/CFT correspondence, this formalism was first found in \cite{Janik:2007wt} and then generalized in \cite{Bajnok:2008bm}, where it was employed to find the spectrum of the Konishi operator at four loops. This result was in perfect agreement with direct field theory calculations \cite{Fiamberti:2008sh,Velizhanin:2008jd}. The L\"uscher formula approach was then extensively used to find anomalous dimensions of the so called twist-$J$ operators. The spectrum of the twist-2 operators was found up to five-loop order \cite{Bajnok:2008qj,Bajnok:2009vm,Lukowski:2009ce} and for twist-3 operators even to six loops \cite{Beccaria:2009eq,Velizhanin:2010cm}. 

The study of the spectrum in the finite volume culminated in the formulation of the Y-system and the Thermodynamic Bethe Ansatz (TBA) equations \cite{Arutyunov:2009zu,Gromov:2009tv,Gromov:2009bc,Bombardelli:2009ns,Arutyunov:2009ur}. Whereas L\"uscher's approach is manifestly perturbative in nature, these equations are believed to describe the full planar spectrum of $\mathcal{N}=4$ super Yang-Mills theory. Both to leading and subleading order in perturbation theory agreement was found with L\"uscher's approach, establishing the compatibility of these formalisms \cite{Arutyunov:2010gb,Balog:2010xa,Balog:2010vf}.


Recently, the techniques used to study finite-size correction in $\mathcal{N}=4$ SYM have been generalized to theories with less supersymmetry, namely, the so-called $\beta$-deformed theories. This was done by introducing twisted transfer matrices \cite{Gromov:2010dy,Arutyunov:2010gu} and by considering deformations of the S-matrix \cite{Ahn:2010yv,Ahn:2010ws}. These methods proved to be succesful as they correctly reproduced wrapping energy corrections that were computed in $\beta$-deformed SYM \cite{Fiamberti:2008sm,Fiamberti:2008sn}. 

The method based on twisting transfer matrices also potentially allows for studying the more general $\gamma$-deformations and orbifold models based on the $\ads$ superstring \cite{Gromov:2010dy,Arutyunov:2010gu}. These developments open up new avenues along which finite size corrections can be studied. Mainly because, in these deformed theories, wrapping corrections appear generically at lower loop orders than in $\mathcal{N}=4$ SYM. In the case of $\beta$-deformed theories wrapping corrections were found to start to contribute one loop order lower and in orbifold models potentially even two orders lower \cite{Arutyunov:2010gu}.

In this paper we study finite-size effects in $\beta$-deformed theories using L\"uscher's approach. We focus on twist operators in the $\alg{sl}(2)$ sector, for which unfortunately little explicit field theory data is known. We consider twist-2 and twist-3 operators. For both families of operators we study the leading order (LO) wrapping correction and we find explicit formulae describing these in terms of harmonic sums. For twist-2 operators we then study the next-to-leading order (NLO) wrapping correction. Our formula agrees with the results recently obtained in \cite{Beccaria:2010kd}, where the authors studied the large-$M$ behaviour of the LO wrapping corrections and explicitly found the first few values for small spins $M$. For general $M$ the leading finite-size correction takes the form
\begin{align}\label{eqn;ENLtwist2intro}
\frac{E_{LO}(M)}{g^6 \sin^2 (2\pi\beta)} =& S_1(M)\frac{S_2(M-1)-S_{-2}(M-1)-S_{-2}(M+1)-S_2(M+1)}{M(M+1)}.
\end{align}
It would be very interesting to have this results confirmed by a direct field theoretic computations. The large spin behavior of (\ref{eqn;ENLtwist2intro}) can be easily read off and it is proportional to $\log M/M^2$ which does not affect the cusp anomalous dimension as suspected. In the following we will also confirm that this result respects reciprocity and compare it with the leading order BFKL equation \cite{Lipatov:1976zz,Kuraev:1977fs,Balitsky:1978ic}. Unfortunately, we were not able to find a closed formula for the NLO wrapping correction to twist-2 operators.

For twist-3 operators we also found a closed formula for the leading wrapping correction
\begin{align}\label{eqn;ENLtwist3intro}
\frac{E_{LO}^{\mbox{\tiny twist-3}}}{g^8 \sin^2 (3\pi \beta)} = \frac{S_1(M/2)}{M+1}\left[\frac{5}{2}\zeta(5)
- S_2(M/2)\zeta(3) + \frac{1}{4}S_5(M/2)-\frac{1}{2}S_{2,3}(M/2)\right].
\end{align} 
Again, one sees that in this case the cusp anomalous dimension is not affected by the wrapping corrections and the result is reciprocity respecting. We will show that both leading order corrections (\ref{eqn;ENLtwist2intro}) and (\ref{eqn;ENLtwist3intro}) are of maximal transcendentality.

This paper is organized as follows. First we will give a brief discussion on twist-$J$ operators and set notation and give definitions. After this we will discuss the twisted transfer matrices that are needed to describe finite-size effects in $\beta$-deformed theories and we will explicitly give the corresponding Y-functions. In section 4 we will then proceed with the computation of the leading order wrapping corrections and study their properties. Subsequently, we continue with a discussion of the next-to-leading order wrapping correction. We end with some conclusions. 

\section{Definitions}

In this section we will introduce the basic notions that will be used in the remainder of this paper. In particular we will define twist-$J$ operators and describe properties of their anomalous dimensions at weak coupling. In the following we will be interested only in the twist-2 and twist-3 operators and we will restrict to these cases whenever necessary. 

\subsection{Twist-$J$ operators}

We will investigate twist-$J$ operators which are embedded in the $\mathfrak{sl}(2)$ closed subsector of the $\beta$-deformed $\mathcal{N}=4$ SYM theory. Their highest weight representatives are composed of $J$ scalar fields $\mathcal{Z}$ and an even number $M$ of covariant derivatives $\mathcal{D}$
\begin{equation}
\mathcal{O}_{M,J} = \mathcal{D}^M\mathcal{Z}^J+\ldots\,,
\end{equation} 
where dots refer to different distributions of covariant derivatives over the scalar fields. For given $J$ and $M$ we will be interested only in the state with the lowest anomalous dimension.

The scaling dimension for twist-$J$ operators can be written as the loop expansion
\begin{equation}
\Delta=J+M+\sum_{\ell=1}^\infty \gamma_{2\ell}\,g^{2\ell}
\end{equation}
where $g^2=\frac{\lambda}{4\pi^2}$. For low loop levels, the anomalous dimension can be found from the Bethe equations. For the $\mathfrak{sl}(2)$ sector of the $\beta$-deformed theory, these are exactly the same as in the non-deformed theory and take the form 
\begin{equation}
\left (\frac{x_{k}^{+}}{x_{k}^{-}}\right)^{J}=\prod_{\substack{j=1\\j\neq k}}^{M} 
\frac{x_{k}^{-}-x_{j}^{+}}{x_{k}^{+}-x_{j}^{-}} 
\frac{1-1/x_{k}^{+}x_{j}^{-}}{1-1/x_{k}^{-}x_{j}^{+}}\,\exp(2i\theta(u_{k},u_{j}))\,, 
\quad \prod_{k=1}^M \frac{x_k^+}{x_k^-}=1\,,
\end{equation}
where the parameters $x^\pm(u)$ are given by the Zhukovsky map and $\theta(u_1,u_2)$ is the dressing phase.  Let us focus here on the twist-2 operators - an analogous discussion is valid for twist-3 operators. Solving the Bethe equations for $J=2$ we get the following asymptotic expression for the anomalous dimensions
\begin{align}
\gamma_{\mbox{\tiny asymp}}^{\mbox{\tiny twist-2}} =\, & 2\,S_1 \,g^2-\left[\, S_1\, S_2+\frac{1}{2}\,S_3\right]g^4 + 
\left[ \frac{1}{2}\, S_1\, S_2^2 + \frac{3}{4} \, S_3\, S_2+\right.\nonumber\\ 
& + \left. \frac{1}{4}\, S_1\, S_4 + \frac{5}{8}\, S_5 - \frac{1}{2} \,S_{2,3} + \, S_1\, S_{3,1} + \frac{1}{2}\, S_{4,1} - S_{3,1,1}\right]g^6+ \ldots.
\end{align}
Additionally, starting at three loops, the anomalous dimension will receive wrapping corrections. These can be conveniently described by the  perturbative procedures applied in \cite{Gromov:2010dy,Arutyunov:2010gu}. This behaviour differs from the non-deformed case where the wrapping correction to the twist-2 operators anomalous dimension starts at four-loop order. The reason is that in the non-deformed case, supersymmetry delays the finite-size effects to higher loop order (e.g. operator $\mathcal{O}_{2,2} $ belongs to the same supermultiplet as Konishi which has length $L=4$). The $\beta$-deformation reduces the supersymmetry from $\mathcal{N}=4$ to $\mathcal{N}=1$ and one can no longer find states with length $L=4$ but rather one finds states with lower length in the twist-$2$ operators multiplets. Because leading wrapping corrections are of order $g^{2L}$, this means that they will start a lower loop order in the $\beta$-deformed theory compared to $\mathcal{N}=4$ SYM.

\subsection{Harmonic sums}

Similarly to the $\mathcal{N}=4$ SYM theory, the anomalous dimensions are expressed in terms of transcendental functions: Riemann $\zeta$-functions and nested harmonic sums $S_{a_1,\ldots,a_m}$. Since the former are well-known functions we only define the latter. For one index we have the standard definition of the harmonic sums
\begin{align}
S_a(M) \equiv \sum_{j=1}^M\frac{(\mathrm{sign}~ a)^j}{j^{|a|}}.
\end{align}
while harmonic sums with multiple indices are defined recursively via
\begin{align}
S_{b,a_1,\ldots a_m}(M) \equiv \sum_{j=1}^M S_{a_1,\ldots a_m}(j)\frac{(\mathrm{sign}~ b)^j}{j^{|b|}}.
\end{align}
The degree of transcendentality is given by the argument of $\zeta$-function or for the nested harmonic sum $S_{a_1,\ldots,a_m}$ by
\begin{equation}
|a_1|+\ldots+|a_m|.
\end{equation}
According to the maximal transcendentality principle anomalous dimension at $\ell$-loop order can be written in terms of the functions with transcendentality degree $2\ell-1$.

\subsection{Reciprocity}
The anomalous dimension $\gamma(M)$ is conjectured to obey a powerful constraint known as  the generalized Gribov-Lipatov reciprocity. This constraint, arising in the QCD context, has been presented in~\cite{Dokshitzer:2005bf,Dokshitzer:2006nm} and approached in~\cite{Basso:2006nk} from the point of view of the large $M$ expansion. In particular, in~\cite{Basso:2006nk} such an analysis has been generalised to anomalous dimensions of operators of arbitrary twist-$J$. Reciprocity  has been checked in various multi-loop calculations of weakly coupled ~\cite{Beccaria:2009vt,Forini:2008ky,Beccaria:2007bb,Beccaria:2007pb,Beccaria:2007cn,Beccaria:2009eq} and strongly coupled \cite{Beccaria:2008tg,Beccaria:2010ry} ${\cal N}=4$ gauge theory. 

The reciprocity constraint can be easily expressed in terms of the $P$-function depending on the spin $M$. This function is  in one-to-one correspondence, at least perturbatively, with the anomalous dimension $\gamma(M)$ as follows from~\cite{Dokshitzer:2005bf,Basso:2006nk,Dokshitzer:2006nm}
\be
\label{nonlinear}
\gamma(M) = P\left(M+\textstyle{\frac{1}{2}\gamma(M)}\right).
\ee
The reciprocity condition is a constraint that arises in the large $M$ expansion of $P(M)$, which is expected to take the 
following form
\be
\label{RRJ}
\qquad P(M) = \sum_{n\ge 0} \frac{a_n(\log\,K^2)}{K^{2\,n}}, \qquad K^2=M\,\left(M+1\right),
\ee 
where the $a_n$ are coupling-dependent polynomials. Eq.~(\ref{RRJ}) implies an infinite 
set of constraints on the coefficients of the large $M$ expansion of $P(M)$ organized in a standard $1/M$ 
power series. We see in \eqref{RRJ} the absence of terms of the form 
$1/K^{2n+1}$, odd under $K\to -K$. 

\subsection{Large $M$ asymptotics and BFKL}

Usually there are two additional checks one can make to test the correctness of the obtained result. Firstly, the large $M$ limit of the anomalous dimension for twist-2 operators is related to the cusp anomalous dimension \cite{Korchemsky:1988si,Korchemsky:1992xv}
\begin{equation}\label{cusp.anomalous.dimension}
\lim_{M\to \infty}\gamma(M) =2\gamma_{\mbox{\tiny cusp}}(g)\log M+\ldots\,,
\end{equation} 
where $\gamma_{\mbox{\tiny cusp}}(g)$ can be investigated both from the perturbative side \cite{Bern:2006ew} and from the strong coupling side \cite{Basso:2007wd} with an interpolating answer coming from the BES equation \cite{Beisert:2006ez}. It is known that cusp anomalous dimension in \eqref{cusp.anomalous.dimension} is completely reproduced by the Asymptotic Bethe Ansatz result which leaves us with the conclusion that the wrapping contribution should be subleading in the large $M$ limit.

Additionally, the continuation of the anomalous dimension to the non-physical values of the spin $M=-1$ should be in agreement with the BFKL equation. Unfortunately, the BFKL equation for $\beta$-deformed  theory is not known. However, the leading contribution is expected to be exactly the same as for the non-deformed theory. It comes from the fact that $\beta$-deformation affects only the superpotential in the action, which is not relevant for the leading BFKL ressumation. The analytic continuation of our result should therefore agree with the following expansion coming from BFKL
\begin{align}\label{eqn;BFKL}
 \gamma&=&\left(2+\mathcal{O}(\omega)\right)
\left(\frac{-\,g^2}{\omega}\right) -\left(0+\mathcal{O}(\omega)\right)\,\left(\frac{-\,g^2}{\omega}\right)^2
+\left(0+\mathcal{O}(\omega) \right)\,\left(\frac{-\,g^2}{\omega}\right)^3+\ldots
\end{align}

\section{Twisted Y-functions}

At this point we will turn to the explicit computation of L\"uscher formulae for $\beta$-deformed theory. In this section we will discuss the Y-functions that describe the wrapping corrections for the twist-2,3 operators that we are considering. We will be able to write general expressions for these Y-functions in terms of Baxter polynomials. Before moving on to the details of the twist-2,3 cases separately, let us first focus on some general features of the relevant Y-functions. 

\subsection{Twisted transfer matrix}

The key feature that allows one to describe $\beta$-deformed theory is the notion of a twisted transfer matrix, for more details see \cite{Gromov:2010dy,Arutyunov:2010gu}.  Consider $M$ string theory particles characterized by the rapidities $u_1,\ldots, u_M$. Consider also an auxiliary particle with rapidity $v$ corresponding to a bound state representation $\pi_Q$ of $\su(2|2)$ with bound state number $Q$. Scattering this auxiliary particle through $M$ particles gives rise to a monodromy matrix
$$
\mathbb{T}(v|\vec{u})=\prod_{i=1}^M \mathbb{S}_{ai}(v,u_i)\, .
$$ 
Here $\mathbb{S}_{ai}(v,u_i)$ is the S-matrix which describes scattering of the auxiliary particle with a particle with rapidity $u_i$. As a matrix acting on the auxiliary space, $\mathbb{T}(v|\vec{u})$ satisfies the fundamental commutation relations
$$
\mathbb{S}_{12}(v_1,v_2)\mathbb{T}_1(v_1|\vec{u})\mathbb{T}_2(v_2|\vec{u})=\mathbb{T}_2(v_2|\vec{u})\mathbb{T}_1(v_1|\vec{u})
\mathbb{S}_{12}(v_1,v_2)\, .
$$
We can introduce a twisted transfer matrix 
\begin{align*}
T(v|\vec{u})={\rm Tr}\Big[\pi_Q(g)\, \mathbb{T}(v|\vec{u})\Big] \, , 
\end{align*}
where the element $g$ is called the twist and the trace is taken over the auxiliary space. If $g$ is such that $[\mathbb{S}_{12},g\otimes g]=0$, then the fundamental commutation relations imply that $T(v|\vec{u})$ commute for different values of $v$ and therefore define a set of commuting charges. For the case at hand we are interested in  $g\in {\rm SU(2)}\times {\rm SU(2)}$.
 
It was found that in order to describe $\beta$-deformed theory, the left and right sector have to be twisted differently.  More specifically, for $\alg{sl}(2)$, the left sector remains untwisted while the right sector is twisted with a twist of the form $g=1\otimes K$, with 
\begin{align}
K = \begin{pmatrix}
 e^{2\pi i J\beta} & 0 \\
0 & e^{-2\pi i J\beta} 
\end{pmatrix}.
\end{align}
Note that this twist depends on $J$ and consequently is different for twist-2,3.

Asymptotically, the Y-functions{\footnote{Since we are mainly interested in the asymptotic solution we will omit the superscript $o$ that indicates the asymptotic solution in the rest of the paper.}} are given by the generalized L\"uscher's formula \cite{Bajnok:2008bm}
\begin{align}
Y^{o}_Q(v) = e^{-J\tilde{\mathcal{E}}_Q(v)}T^{l}(v|\vec{u})T^{r}(v|\vec{u})\prod_i S^{Q1_*}_{\alg{sl}(2)}(v,u_i).
\end{align}
Here $\tilde{\mathcal{E}}_Q(v)$ is the energy of a mirror $Q$-particle, $S^{Q1_*}_{\alg{sl}(2)}(v,u_i)$ denotes the S-matrix with arguments in the mirror ($v$) and string regions ($u_i$) and finally $T^{l,r}$ are the left and right twisted transfer matrices. 

For the states from the $\sl(2)$ sector one has that both $T^{l}$ and $T^{r}$ are described by $T_{Q,1}$, which is given by
\bea\label{TS}
T_{Q,1}(v\,|\,\vec{u})&=&1+\prod_{i=1}^{M} \frac{(x^--x^-_i)(1-x^-
x^+_i)}{(x^+-x^-_i)(1-x^+
x^+_i)}\frac{x^+}{x^-}\\
&&\hspace{-1.5cm}-2\cos\alpha\sum_{k=0}^{Q-1}\prod_{i=1}^{M}
\frac{x^+-x^+_i}{x^+-x^-_i}\sqrt{\frac{x^-_i}{x^+_i}}
\left[1-\frac{\frac{2ik}{g}}{v-u_i+\frac{i}{g}(Q-1)}\right]+\sum_{m=\pm}
\sum_{k=1}^{Q-1}\prod_{i=1}^{M}\lambda_m(v,u_i,k)\, . \nonumber
\eea Definitions of various quantities entering the last formula
can be found in appendix A; $\cos{\alpha}$ is a twist of the bosonic
eigenvalues. From the discussion above we have that $\alpha_{l} = 0$ and $\alpha_{r} = 2\pi J\beta$, where $\beta$ is the deformation parameter of the theory.

As we are interested in the leading order wrapping correction, we will only evaluate our Y-function to the lowest order in $g$ in this section. Accordingly, we will denote this lowest order simply by $Y_Q$, leaving the $g$-expansion implicit. It is important to stress that the wrapping correction for twist-2,3 operators starts at 3 and 4 loops respectively. This is one order lower as is the case in $\mathcal{N}=4$ SYM.

\subsection{Twist-2}

The one-loop Bethe roots $u_i$ in the $\alg{sl}(2)$ sector for twist-2 operators can be encoded by the Baxter polynomial \cite{Derkachov:2002tf,Eden:2006rx}
\begin{align}
P_{M}(u)=\, _3F_2\left(-M,M+1,\frac{1-i u}{2};1,1;1\right).
\end{align}
The zeroes of this polynomial give, to lowest order in $g$, the solutions to the Bethe equations in the $\alg{sl}(2)$ sector. 

Similarly to \cite{Bajnok:2008qj}, we find that the Y-function (to lowest order in $g$) $Y_Q(M)$ for $\beta$-deformed theory can be written purely in terms of the Baxter polynomial $P_M$
\begin{align}\label{eqn;Ytwist2}
Y_{Q}(M) = g^6\sin ^2 (2 \pi \beta) \frac{T_Q(M)\tilde{T}_Q(M)}{R_Q(M)}\frac{S_1(M)}{(v^2+Q^2)^2},
\end{align}
where we defined
\begin{align}
T_Q(M) &= \sum _{k=0}^{Q-1} \left[\frac{1}{2 k - Q - i v}-\frac{1}{2 (k+1) - Q - i v}\right] P_M\left(v+i\frac{2k- Q + 1}{2}\right)\label{eqn;Tq}\\
\tilde{T}_Q(M) &= 4  \sum _{k=0}^{Q-1} P_M \left(v - i\frac{2 k-Q+1}{2}\right)\label{eqn;Ttildeq}.
\end{align}
The denominator is given by{\small
\begin{align}
R_Q(M)=P_M\left(v-\frac{i(Q+1)}{2}\right)P_M\left(v-\frac{i(Q-1)}{2}\right)P_M\left(v+\frac{i(Q-1)}{2}\right)P_M\left(v + \frac{i(Q+1)}{2}\right)\label{eqn;Rq}.
\end{align}}
The fact that the Y-functions factor into two different parts $T_Q,\tilde{T}_Q$ is a direct consequence of the different twistings of the left and right sectors. 

Let us remark that the wrapping correction is proportional to $S_1(M)$. We would also like to point out that the complete $\beta$ dependence at this level is simply given by a factor of $\sin ^2(2 \pi \beta)$ in front of the wrapping correction. This automatically ensures that when sending $\beta\rightarrow 0$, the wrapping correction vanishes and our result agrees with the calculations in $\mathcal{N}=4$ SYM. Finally, the $\beta$-dependent correction also vanishes upon sending $\beta\rightarrow \frac{1}{2}$.

\subsection{Twist-3}

Analogously, the one-loop Bethe roots in the $\alg{sl}(2)$ sector for twist-3 operators are encoded by the Baxter polynomial \cite{Kotikov:2007cy}
\begin{align}
\tilde{P}_{M}(u)= \, _4F_3\left(\frac{M}{2}+1,-\frac{M}{2},\frac{1-iu}{2},\frac{1+iu}{2};1,1,1;1\right)
\end{align}
Again, the zeroes of this polynomial give, to lowest order in $g$, the solutions to the Bethe equations.

The Y-function for twist-3 can be written exactly in the same way as for twist-2 operators in terms of the Baxter polynomial $P_M$, namely
\begin{align}
Y_{Q}(M) = g^8 \sin ^2 (3\pi \beta) \frac{T_Q(M)\tilde{T}_Q(M)}{R_Q(M)}\frac{S_1(\frac{M}{2})}{(v^2+Q^2)^3},
\end{align}
where $T_Q,\tilde{T}_Q$ and $R_{Q}$ are defined as in (\ref{eqn;Tq}),(\ref{eqn;Ttildeq}) and (\ref{eqn;Rq}) respectively, but with $P_M$ replaced by the twist-3 polynomial $\tilde{P}_M$.

This time the wrapping correction is proportional to $S_1(\frac{M}{2})$. Again the $\beta$ dependence at this level is simply given by a $\beta$-dependent prefactor that vanishes for $\beta\rightarrow0$ and for $\beta\rightarrow \frac{1}{3},\frac{2}{3}$ {\footnote{Our results exhibit the property that for special values of $\beta=\frac{n}{J}$ the leading wrapping correction vanishes as was the case in \cite{Arutyunov:2010gu}. The result is also manifestly invariant under the shift $\beta\rightarrow \beta+\frac{n}{J}$, cf. \cite{Gunnesson:2009nn}.}}.

\section{Wrapping correction}

In this section we present the wrapping corrections to the twist-2,3 operators. We find that both wrapping corrections can be written in terms of harmonic sums. However, in contradistinction to $\mathcal{N}=4$ SYM, the arguments of the different sums are shifted.

The energy of an $M$-particle state from the $\sl(2)$-sector is given by \cite{Gromov:2009tv,Gromov:2009bc,Bombardelli:2009ns,Arutyunov:2009ur}
\begin{align}
\label{FSE} E =J+\sum_{i=1}^M{\cal E}(p_i)
-\frac{1}{2\pi}\sum_{Q=1}^{\infty}\int {\rm d}v
\frac{d\tilde{p}^Q}{dv}\log(1+Y_{Q}(v)).
\end{align}
Here the integration runs over a real rapidity line of the mirror theory and $\tilde{p}^Q$ are momenta of the mirror $Q$-particles. Moreover, ${\cal E}(p)$ is the asymptotic energy of a string theory particle with momentum $p$, given by \bea {\cal E}(p)=\sqrt{1+4g^2\sin^2\frac{p}{2}}\, . \eea The last term in the formula (\ref{FSE}) can be understood as the finite-size correction to the asymptotic, {\it i.e.} large $J$, dispersion relation.

When expanding energy and momentum for small $g$ around the asymptotic solutions $Y^{o}_Q$, we obtain the leading order
corrections to the energy 
\begin{align}\label{eqn;Luscher} E_{\rm LO} =
-\frac{1}{2\pi}\sum_{Q=1}^{\infty}\int {\rm d}v\,  Y_{Q}(v)\, .
\end{align}
By using this formula and our explicit expressions for the Y-functions derived in the previous section, it is now straightforward to compute the leading order finite-size correction to the energy.

\subsection{Twist-2}

We use formula (\ref{eqn;Ytwist2}) for the Y-function to calculate (\ref{eqn;Luscher}) for operators with an even number of particles. We have computed the wrapping correction for $M=2,4,\ldots,90$ {\footnote{Our findings agree with the results listed in formula (3.20) of \cite{Beccaria:2010kd}.}}. 

It is quickly seen that the resulting $E_{LO}(M)$ cannot be written purely in terms of harmonic sums $S_{a_1,a_2,\ldots}(M)$. However, we found that it is possible to write the wrapping correction as a combination of harmonic sums with shifted arguments:
\begin{align}
\frac{E^{\mbox{\tiny twist-2}}_{LO}(M)}{g^6 \sin^2(2\pi\beta)} =& \frac{2S_1(M)}{M(M+1)}\left[(A + B +2) S_{-2}(M) - (A - B )S_2(M) - A  S_{-2}(M-1)+\right.\nonumber\\
&\left.\qquad\qquad\qquad\ \ \ +A  S_2(M-1) - B  S_{-2}(M+1) - B S_2(M+1)\right]
\end{align}
where $A,B$ are arbitrary constants. The combination of terms proportional to $A,B$ vanishes for even values of $M$. In other words, for even $M$ the above result coincides with
\begin{align}\label{eqn;SimpleTwist2}
E^{\mbox{\tiny twist-2}}_{LO}(M) =&~ 4g^6 \sin^2 (2\pi\beta)\frac{S_1(M)S_{-2}(M)}{M(M+1)}
\end{align}
It is directly seen that formula (\ref{eqn;SimpleTwist2}) is reciprocity respecting. It is also useful to notice that the denominator $M(M+1)$ can be written as
\begin{align}
\frac{1}{M(M+1)} = (S_1(M+1)-S_1(M))(S_1(M)-S_1(M-1)).
\end{align}
This indicates that the degree of transcidentality is 5, and as such, the wrapping correction in $\beta$-deformed theory obeys the principle of maximum transcendentality\footnote{Alternatively, one can write $\frac{1}{M(M+1)}=\frac{1}{M}-\frac{1}{M+1}$ and assign a transcendentality degree 1 to $\sin^2(2\pi\beta)$.}. It is also easy to check that, in the limit of large spin $M$, formula \eqref{eqn;SimpleTwist2} behaves like 
\begin{align}
E^{\mbox{\tiny twist-2}}_{LO} \sim \frac{\log M}{M^2}+\ldots\,.
\end{align}

The coefficients $A,B$ can subsequently be fixed by considering the BFKL equation and reciprocity. Analytically continuing $E^{\mbox{\tiny twist-2}}_{LO}(M)$ to $M=-1$ reveals the following pole structure
\begin{align}
\frac{2 (B+1)}{\omega ^4}+\frac{2 (B+1)}{\omega ^3}+\frac{2 (B+A+1)-\frac{1}{6} \pi ^2 (5 B+8)}{\omega^2}+\mathcal{O}\left(\frac{1}{\omega }\right),
\end{align}
where $\omega = M+1$. According to the BFKL prediction for the undeformed theory \eqref{eqn;BFKL}, the pole structure at this level should start at $\omega^{-2}$. This can uniquely be achieved by setting
\begin{align}
B=-1.
\end{align}
Moreover, one then finds that the reciprocity condition \eqref{RRJ} is satisfied only for $A=B$ and hence we arrive at the following final result for twist-2 operators 
\begin{align}\label{eqn;ELOtwist2}
\frac{E^{\mbox{\tiny twist-2}}_{LO}(M)}{g^6 \sin^2\pi\beta} =& S_1(M)\frac{S_2(M-1)-S_{-2}(M-1)-S_{-2}(M+1)-S_2(M+1)}{M(M+1)}.
\end{align}
Of course, the above form is fixed by requiring reciprocity and compatibility with BFKL, which are most likely valid for $\beta$-deformed SYM, but it would be interesting to find field theoretic evidence for this.

\subsection{Twist-3}

We explicitly computed the leading contribution to twist-3 operators for $M=2,4,\ldots,64$. In this case, we found the following form of the wrapping correction
\begin{align}\label{eqn;ELOtwist3}
\frac{E^{\mbox{\tiny twist-3}}_{LO}}{g^8 \sin^2 (3\pi \beta)} = \frac{S_1(M/2)}{M+1}\left[\frac{5}{2}\zeta(5)
- S_2(M/2)\zeta(3) + \frac{1}{4}S_5(M/2)-\frac{1}{2}S_{2,3}(M/2)\right].
\end{align} 
It is worthwhile to notice that all the indices of the harmonic sums are positive and depend on $M/2$, similar to $\mathcal{N}=4$ SYM \cite{Beccaria:2007pb,Beccaria:2009eq,Velizhanin:2010cm}. The maximum transcendentality principle and reciprocity are once again respected.

\section{NLO wrapping corrections}

To describe the next-to-leading order wrapping correction all quantities in (\ref{FSE}) must be carefully expanded to one order higher in $g^2$. In what follows we will first discuss the expansion of the different terms separately and then give the resulting NLO wrapping correction to the energy.

\subsection{Different contributions}

First we consider the term in (\ref{FSE}) that is not related to the Y-function
\begin{align}
\frac{d\tilde{p}^Q}{dv} = 1 + 2g^2\frac{v^2-Q^2}{(v^2+Q^2)^2}
+\mathcal{O}(g^4).
\end{align}
Next we focus on the term $\log (1+Y_Q(v|\hat{u}))$. This term depends both implicitly (through the Bethe root $\hat{u} = u + g^2\delta u$) and explicitly on $g$. Expanding this gives
\begin{align}
\log (1+Y_Q(v|\hat{u})) = Y^{\rm LO}_Q(v|u) + Y^{\rm LO,2}_Q(v|u)
+ Y^{\rm NLO}_Q(v|u) + \mathcal{O}(g^{10}),
\end{align}
where $Y^{\rm LO}_Q(v|u)$ is of order $g^{6}$ and $Y^{\rm LO ,2}_Q(v|u), Y^{\rm NLO}_Q(v|u)$ are of order $g^{8}$. Here $Y^{\rm LO,2}_Q(v|u)$ is obtained by expanding the Bethe root $\hat{u} = u + g^2\delta u$ and is  expressed as
\begin{align}
Y^{\rm LO,2}_Q(v|u) &\equiv g^2\,\sum_i \partial_{u_i} Y^{\rm LO}_Q(v|u)\,
\delta u_i .
\end{align}
Subsequently, we turn our attention to $Y^{\rm NLO}_Q(v|u)$. This function is easily computed by expanding (\ref{TS}). The untwisted transfer matrix coincides with the one used for $\mathcal{N}=4$ SYM, so let us focus on the twisted transfer matrix. This transfer matrix admits the following expansion
\begin{align}
T^{NLO}_{Q}(\alpha) = \frac{2g^2S_1(M)}{Q-iv}T^{LO}_{Q}(\alpha) + T^{LO}_Q(\alpha=0).
\end{align}
The second term in the above expansion automatically ensures that upon sending $\beta\rightarrow 0$, our result reproduces the correct $\mathcal{N}=4$ SYM result. In $Y^{\rm NLO}_Q(v|u)$ the expansion of the scalar factor $S_0$ of the S-matrix also has to be taken into account. Its explicit small $g$ expansion is given in Appendix A.

\smallskip

Finally, we consider the term involving the dispersion relation
$\mathcal{E}(p)$ from eq.(\ref{FSE}). Since the momentum also
receives a wrapping correction, {\it i.e.}
\begin{align}
p \rightarrow p + g^{6}\delta p,
\end{align}
the asymptotic energy ${\cal E}(p)$ also gets corrected
\begin{align}
\mathcal{E}(p) &= \sqrt{1+4 g^2\sin^2\frac{p + \delta p}{2}} =
\sqrt{1+4 g^2\sin^2\frac{p}{2}} + g^{8}\sin p\, \delta p  +
\mathcal{O}(g^{10}).
\end{align}
The correction $\delta p$ can be computed from the Bethe equations \cite{Bajnok:2009vm}. Define the following function
\begin{align}
BAE_k =-\left(\frac{u_k + i}{u_k-i}\right)^2\prod _{j=1}^{M} \frac{u_k-u_j + 2 i}{u_k-u_j-2 i}.
\end{align}
The correction to the momentum $\delta p$ is then described by the following set of equations
\begin{align}
\sum_i \frac{\partial BAE_k}{\partial p_i}\delta p_i = \Phi_k, \qquad k=1,\ldots M,
\end{align}
where momentum and rapidity are related via $u = \cot\frac{p}{2}$ and $\Phi_k$ is given by
\begin{align}
\Phi_k = \sum_{Q=1}^{\infty}\int dv \frac{d\tilde{p}^Q}{dv} \frac{g^6}{(v^2+Q^2)^2}\mathrm{str}_a\left[\mathbb{S}_{a1}(v,u_1)\ldots\partial_v\mathbb{S}_{ak}(v,u_k)\ldots \mathbb{S}_{aM}(v,u_M)\right].
\end{align}
From the explicit form of $\Phi_k$ it is readily seen that $\delta p \sim \sin^2(2\pi\beta)$. Hence, the correction to the Bethe roots at this order is purely an effect of the $\beta$-deformation. It would be interesting to see if this correction can also be derived from the thermodynamic Bethe Ansatz approach, along the lines of \cite{Arutyunov:2010gb,Balog:2010xa,Balog:2010vf}.

Concluding, we see that the NLO wrapping correction to the energy splits into two pieces
\begin{align}
E_{NLO}(\beta) = E^{\mathcal{N}=4}_{LO} + \sin^2(2\pi\beta) ~ E^{(\beta)}_{NLO}.
\end{align}
The correction $E^{\mathcal{N}=4}_{LO}$ has been computed in \cite{Bajnok:2008qj} and consequently we will only focus on $E^{(\beta)}_{NLO}$.

\subsection{Results}

By carefully computing the different terms discussed above, the first few next-to-leading order wrapping corrections to the energy were found to be
\begin{center}
\begin{tabular}{|c|c|}\hline
M & $E^{(\beta)}_{NLO}/g^8$ \\
\hline \hline
2 & $\frac{3}{2} \zeta (3)-\frac{87}{32}$\\ \hline 
4 & $\frac{125}{144}\zeta(3)-\frac{634475}{497664}$\\ \hline 
6 & $\frac{343}{600}\zeta(3)-\frac{195848051}{259200000}$\\ \hline
8 & $\frac{579121}{1411200}\zeta(3)-\frac{849922576886413}{1672847769600000}$\\ \hline
10 & $\frac{4952651}{15876000}\zeta(3)-\frac{187527445351389407}{508127510016000000}$\\ \hline  
12 & $\frac{569200957}{2305195200}\zeta(3)-\frac{27728787085943160324263}{98201332338704179200000}$\\ \hline
14 & $\frac{1372958223289}{6817614804000}\zeta(3)-\frac{143089155610576965157748149667}{638075677540603589134848000000}$\\ \hline
16 & $\frac{349224691793}{2077749273600}\zeta(3)-\frac{284957306041515539714929299089}{1555689270956138274462105600000}$\\ \hline  
18 & $\frac{10723982580979}{75058692508800}\zeta(3)-\frac{42226000076368077644613444015013001}{276107037648996202745367733862400000}$\\ \hline
20 & $\frac{623512460093645}{5057955092526336}\zeta (3)-\frac{16584475828775376357486983395513628316781}{127618093832253145076032429417138225152000}$\\\hline
\end{tabular}
\end{center}
The $\zeta(3)$ dependent part can easily be written in closed form as
\begin{align} 
4\frac{(S_1(M))^2}{M(M+1)}\zeta(3).
\end{align}
The rational part, however, is more involved and so far we have not been able to fix its form in terms of harmonic sums. However, a careful examination of the results seems to indicate the following structure
\begin{align}
\frac{S_{transc=5}}{M(M+1)},
\end{align}
where $S_{transc=5}$ stands for a combination of harmonic sums of transcendentality degree 5, possibly with shifted arguments. Again one can notice that for large $M$ this correction vanishes. 

\section{Conclusions}

In this paper we have derived the finite-size corrections for twist-2,3 operators in $\beta$-deformed SYM theory, cf. equations (\ref{eqn;ELOtwist2}) and (\ref{eqn;ELOtwist3}). The obtained results respect the principle of maximum transcendentality as well as reciprocity. Both wrapping corrections go to zero in the limit $M\rightarrow\infty$.

For twist-2 we also studied the pole structure and compared it against leading BFKL. It would be interesting to find the next-to-leading prediction and compare it with our result
\begin{align}
-\frac{2+3\zeta(2)}{\omega^2}g^6.
\end{align}
Furthermore, it also would be important to find a closed formula for the NLO wrapping corrections discussed in section 5, which would give more insight on the structure of finite-size corrections in $\beta$-deformed theories.

One of the nice features of the $\beta$-deformed theory is that wrapping corrections appear at lower loop level than in the undeformed theory, which makes them more accessable and gives hope that double wrapping is within reach. A better understanding of this would give valuable insights in the precise nature of the integrable structures of $\beta$-deformed theories. It would be also interesting to study more general deformations ($\gamma$-deformations) that might exhibit novel structures in the finite-size spectrum.

\section*{Acknowledgments}

We are grateful to G. Arutyunov, N. Beisert, V. Forini, G. Korchemsky,  A. Kotikov, L. Lipatov, C. Sieg and S. van Tongeren for useful discussions and comments. T.~{\L}ukowski is supported by a DFG grant in the framework of the SFB 647 {\it ``Raum - Zeit - Materie. Analytische und Geometrische Strukturen''} .

\section{Appendices}

\subsection{Appendix A: Twisted transfer matrix}
The eigenvalue of the twisted transfer matrix for an
anti-symmetric bound state representation with the bound state
number $Q$ is given by the following formula, generalizing the
result of \cite{Arutyunov:2009iq}
\begin{eqnarray}\label{eqn;FullEignvalue}
&&T_{Q,1}(v\,|\,\vec{u})=\prod_{i=1}^{K^{\rm{II}}}{\textstyle{\frac{y_i-x^-}
{y_i-x^+}\sqrt{\frac{x^+}{x^-}}
\, +}}\\
&&
{\textstyle{+}}\prod_{i=1}^{K^{\rm{II}}}{\textstyle{\frac{y_i-x^-}{y_i-x
^+}\sqrt{\frac{x^+}{x^-}}\left[
\frac{x^++\frac{1}{x^+}-y_i-\frac{1}{y_i}}{x^++\frac{1}{x^+}-y_i-\frac{1
}{y_i}-\frac{2i Q}{g}}\right]}}\prod_{i=1}^{K^{\rm{I}}}
{\textstyle{\left[\frac{(x^--x^-_i)(1-x^-
x^+_i)}{(x^+-x^-_i)(1-x^+
x^+_i)}\frac{x^+}{x^-}  \right]}}\nonumber\\
&&{\textstyle{+}}
\sum_{k=1}^{Q-1}\prod_{i=1}^{K^{\rm{II}}}{\textstyle{\frac{y_i-x^-}{y_i-
x^+}\sqrt{\frac{x^+}{x^-}}
\left[\frac{x^++\frac{1}{x^+}-y_i-\frac{1}{y_i}}{x^++\frac{1}{x^+}-y_i-\frac{1}{y_i}-\frac{2ik}{g}}\right]}}
\left\{\prod_{i=1}^{K^{\rm{I}}}{\textstyle{\lambda_+(v,u_i,k)+}}\right.\left.\prod_{i=1}^{K^{\rm{I}}}{\textstyle{\lambda_-(v,u_i,k)}}\right\}\nonumber\\
&&\quad -\sum_{k=0}^{Q-1}\prod_{i=1}^{K^{\rm{II}}}
{\textstyle{\frac{y_i-x^-}{y_i-x^+}\sqrt{\frac{x^+}{x^-}}\left[\frac{x^+
-\frac{1}{x^+}-y_i-\frac{1}{y_i}}
{x^+-\frac{1}{x^+}-y_i-\frac{1}{y_i}-\frac{2ik}{g}}\right]}}\prod_{i=1}^
{K^{\rm{I}}}{\textstyle{\frac{x^+-x^+_i}{x^+-x^-_i}\sqrt{\frac{x^-_i}{x^
+_i}} \left[1-\frac{\frac{2ik}{g}}{v-u_i+\frac{i}{g}(Q-1)
}\right]}}\times\nonumber\\
&&\quad\times
\left\{e^{i\alpha}\prod_{i=1}^{K^{\rm{III}}}{\textstyle{\frac{w_i-x^+-\frac{1}{x^+}
+\frac{i(2k-1)}{g}}{w_i-x^+-\frac{1}{x^+}+\frac{i(2k+1)}{g}}+ }}
e^{-i\alpha}\prod_{i=1}^{K^{\rm{II}}}{\textstyle{\frac{y_i+\frac{1}{y_i}-x^+-\frac
{1}{x^+}+\frac{2ik}{g}}{y_i+\frac{1}{y_i}-x^+-\frac{1}{x^+}+\frac{2i(k+1
)}{g}}}}\prod_{i=1}^{K^{\rm{III}}}{\textstyle{\frac{w_i-x^+-\frac{1}{x^+
}+\frac{i(2k+3)}{g}}{w_i-x^+-\frac{1}{x^+}+\frac{i(2k+1)}{g}}}}\right\}.
\nonumber
\end{eqnarray}
Here the twist $e^{i\alpha}$  enters only the last line.
Eigenvalues are parametrized by solutions of the auxiliary Bethe
equations:
\begin{eqnarray}
\label{bennote}
\prod_{i=1}^{K^{\rm{I}}}\frac{y_k-x^-_i}{y_k-x^+_i}\sqrt{\frac{x^+_i}{x^
-_i}}&=&e^{i\alpha}
\prod_{i=1}^{K^{\rm{III}}}\frac{w_i-y_k-\frac{1}{y_k}-\frac{i}{g}}{w_i-y
_k-\frac{1}{y_k}+\frac{i}{g}},\\
\prod_{i=1}^{K^{\rm{II}}}\frac{w_k-y_i-\frac{1}{y_i}+\frac{i}{g}}{w_k-y_
i-\frac{1}{y_i}-\frac{i}{g}} &=& e^{2i\alpha}\prod_{i=1,i\neq
k}^{K^{\rm{III}}}\frac{w_k-w_i+\frac{2i}{g}}{w_k-w_i-\frac{2i}{g}}.\nonumber
\end{eqnarray}
In the formulas above the variable
$$
v=x^++\frac{1}{x^+}-\frac{i}{g}Q=x^-+\frac{1}{x^-}+\frac{i}{g}Q\,
$$
takes values in the mirror theory rapidity plane, i.e. $x^\pm = x(v
\pm {i\ov g}Q)$ where $x(v)$ is the mirror theory $x$-function. As
was mentioned above, $u_j$ take values in string theory $u$-plane,
and therefore $x_j^\pm = x_s(u_j \pm {i\ov g})$ where $x_s(u)$ is
the string theory $x$-function. These two functions are given by
\begin{align}
&x(u) = \frac{1}{2}(u-i\sqrt{4-u^2}), && x_s(u) = \frac{u}{2}(u+\sqrt{1-\frac{4}{u^2}}).
\end{align}
Finally, the quantities {\small
$\lambda_{\pm}$ are
\begin{eqnarray}\nonumber \hspace{-1cm}
\lambda_\pm(v,u_i,k)&=&\frac{1}{2}\left[1-\frac{(x^-_ix^+-1)
  (x^+-x^+_i)}{(x^-_i-x^+)
  (x^+x^+_i-1)}+\frac{2ik}{g}\frac{x^+
  (x^-_i+x^+_i)}{(x^-_i-x^+)
  (x^+x^+_i-1)}\right.\\ \label{eqn;lambda-pm}
&&~~~~~~~~~~~~\left.\pm\frac{i x^+
  (x^-_i-x^+_i)}{(x^-_i-x^+)
 (x^+x^+_i-1)}\sqrt{4-\left(v-\frac{i(2k-Q)}{g}\right)^2}\right]\,
 .
\end{eqnarray}
}

The S-matrix in the string-mirror region $S_{\sl(2)}^{1_*Q}$ is
found in \cite{Arutyunov:2009kf} (see also \cite{Bajnok:2009vm}) and it has
the following weak-coupling expansion
$$
S_{\sl(2)}^{1_*Q}(u,v)=S_0(u,v)+g^2 S_{2}(u,v)+\ldots \, ,
$$
where {\small
\begin{align}
S_0(u,v) = -\frac{\big[(v-u)^2+(Q+1)^2\big]\big[Q-1 + i (v-u)
)\big]}{(u-i)^2 \big[Q-1-i( v-u)\big]}.
\end{align}
} and {\small
\begin{align}
S_2(u,v)& =-S_0(v,u)
\frac{2\big[2Q(u-i)+(u+i)(v^2+Q^2+2v(u-i))\big]}{(v^2+Q^2)
(1+u^2)}+\\
\nonumber &
\frac{S_0(v,u)}{1+u^2}\Big[4\gamma+\psi\left(1+\frac{Q+iv}{2}\right)+\psi\left(1-\frac{Q+iv}{2}\right)+\psi\left(1+\frac{Q-iv}{2}\right)+
\psi\left(1-\frac{Q-iv}{2}\right)\Big]\, .
\end{align}
}These expressions are enough to build up the two leading terms in
the weak-coupling expansion of the asymptotic function $Y^o_Q$.

\bibliographystyle{JHEP}
\bibliography{LitTwist}

\end{document}